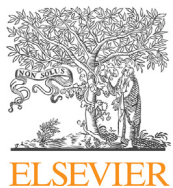
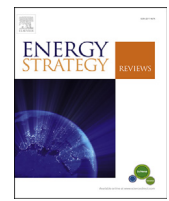

# Opening the black box of energy modelling: Strategies and lessons learned

Stefan Pfenninger [a,*], Lion Hirth [b,c,d], Ingmar Schlecht [e], Eva Schmid [f,g], Frauke Wiese [h], Tom Brown [i], Chris Davis [j], Matthew Gidden [k], Heidi Heinrichs [l], Clara Heuberger [m], Simon Hilpert [n], Uwe Krien [o], Carsten Matke [p], Arjuna Nebel [q], Robbie Morrison [r], Berit Müller [o], Guido Pleßmann [o], Matthias Reeg [s], Jörn C. Richstein [t], Abhishek Shivakumar [u], Iain Staffell [m], Tim Tröndle [v], Clemens Wingenbach [n]

[a] ETH Zürich, Switzerland
[b] Neon Neue Energieökonomik GmbH, Berlin, Germany
[c] Hertie School of Governance, Berlin, Germany
[d] Mercator Research Institute on Global Commons and Climate Change (MCC), Berlin, Germany
[e] University of Basel, Switzerland
[f] GERMANWATCH, Berlin, Germany
[g] PIK, Potsdam, Germany
[h] Danish Technical University, Denmark
[i] FIAS, University of Frankfurt, Germany
[j] University of Groningen, The Netherlands
[k] IIASA, Laxenburg, Austria
[l] Institute of Energy and Climate Research, FZ Jülich, Germany
[m] Imperial College London, UK
[n] University of Flensburg, Germany
[o] Reiner Lemoine Institute, Germany
[p] DLR Institute of Networked Energy Systems, Oldenburg, Germany
[q] Wuppertal Institute, Germany
[r] Energy Consultant, Berlin, Germany
[s] DLR, Germany
[t] DIW, Berlin, Germany
[u] KTH Royal Institute of Technology, Stockholm, Sweden
[v] Cambridge University, UK

**ABSTRACT**

The global energy system is undergoing a major transition, and in energy planning and decision-making across governments, industry and academia, models play a crucial role. Because of their policy relevance and contested nature, the transparency and open availability of energy models and data are of particular importance. Here we provide a practical how-to guide based on the collective experience of members of the Open Energy Modelling Initiative (Openmod). We discuss key steps to consider when opening code and data, including determining intellectual property ownership, choosing a licence and appropriate modelling languages, distributing code and data, and providing support and building communities. After illustrating these decisions with examples and lessons learned from the community, we conclude that even though individual researchers' choices are important, institutional changes are still also necessary for more openness and transparency in energy research.

## 1. Introduction

The history of energy system planning is primarily closed and proprietary, having been pursued by research institutions,

* Corresponding author.
E-mail address: stefan.pfenninger@usys.ethz.ch (S. Pfenninger).

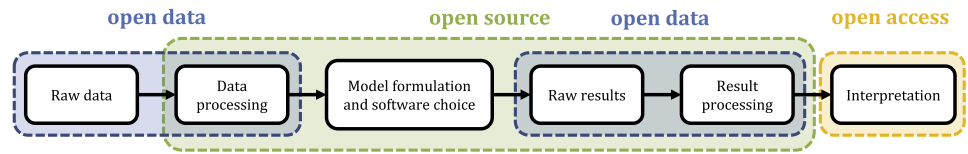

**Fig. 1.** Open data, open source, and open access in relation to the energy modelling process.

government agencies, and large, vertically-integrated utilities that were under no obligation to reveal their modelling assumptions or methodologies. This may have been acceptable in a conventional energy system with only a few players, but the requirements on energy system planning are changing significantly driven by the advent of liberalised, regulated markets and the need for deep reductions in greenhouse gas emissions [1]. In addition, the rapid deployment of wind and photovoltaics (PV) and growing importance of energy storage require models with sufficient spatial and temporal resolution to accurately assess these new technologies.

Open energy modelling is desirable for many reasons [2]. First, open code and data improve scientific quality if they lead to more transparency and reproducibility, and thus permit effective collaboration across the science-policy boundary. This is particularly important in energy policy, an urgent and highly contested topic. More transparent modelling is desirable from a regulatory and political perspective, as opening up decision processes and the reasoning behind them may lessen public opposition to new legislation and infrastructure. By reducing parallel efforts and allowing researchers to collaborate and share the burden of developing and maintaining large codebases and datasets, openness also enables increased productivity. We believe research funded with public money should be freely available to the public.

The open code, open data and open science movements are only slowly leading to models and data being opened up [3]. In order to classify as "open", the data or model code needs to be both accessible and legally usable. Hence, we pragmatically define open code, open data, and open models as artefacts that are available under commonly used licences, which allows re-use without undue restrictions.[1] The history of open energy modelling can be traced to the release of the model Balmorel in March 2001 [5] (albeit without a licence until 2017), followed by early attempts with a now abandoned GPL-licensed model called deeco in 2004 [6]. After a long pause, the release of the modelling frameworks OSeMOSYS in 2009 and TEMOA in 2010 [7,8] has spearheaded several dozen open projects., In 2014, a group of modellers founded the Open Energy Modelling Initiative[2] in order to better coordinate the further development of open models and data. Nevertheless, the mainstream approach to energy modelling is often still proprietary and opaque, even where it directly feeds into policy [9]. Underlying reasons are manifold; however, commercial sensitivity, lock-in to proprietary models, lack of awareness, institutional and personal inertia, and fear of losing competitive advantage are some factors at work in academic research [2]. In addition, the boundary between academic research and commercial consulting is blurred in the energy modelling field, so different actors have different aims and incentives (e.g., selling data or software to customers, or publishing policy-relevant results).

Based on our experience, this article addresses what we perceive to be the crucial factors limiting openness of energy data and models: the lack of practical knowledge as well as personal and institutional inertia. It is not intended as a review of energy modelling methods or tools, but a how-to guide for researchers considering to open up their model code and data. The article proceeds as follows. First, we briefly introduce the key steps in energy modelling and how they link to aspects of openness. We then walk through the practical steps energy modellers must think about and choices they must make when deciding to go open. Finally, we describe four examples that provide further context for these key choices, before concluding with the most important challenges that remain to be overcome in the institutions that shape the research landscape.

## 2. The energy modelling process

The kinds of models we discuss here can be summarized by two overarching terms: energy system models and power system models. The former group includes in particular long-term planning models that generate deployment scenarios over decades into the future, and has recently seen a renaissance with an increased focus on variable renewable generation. The latter group has a clear focus on electricity and often includes more detailed power grid models also used in an operational setting by utility companies. Due to the increasing importance of electricity as a clean energy carrier, power system models and energy system models are increasingly converging.

In this section we outline how data and code relate to the energy modelling process, introducing the concepts for the following sections. We limit our discussion to open data and open source code. The process of publishing results in the literature and the conditions under which the resulting articles are distributed go beyond the scope of this article (see Fig. 1).

### 2.1. Data

Data is both an input and output of the modelling process. Raw data in the energy field is spread widely and of varying quality, coming from academic sources, non-governmental bodies, markets, individuals and commercial entities. An obvious impediment to openness is the widespread use of non-disclosure agreements under which commercially sensitive data may be shared. A less obvious impediment is that in many cases, no explicit licence is attached to input data. Contrary to common practice, this does not imply the legal permission to use and share data, as discussed below. This is of crucial importance since the degree of openness and the licensing conditions of input data influence the degree to which a model based on them can be made open.

Raw data from various sources and in different formats must first be processed to become accessible to a model. Three kinds of data processing are usually necessary: (1) time series data: creating simulated or processing measured intermittent renewable generation and demand data, (2) geographic data (e.g. installed generation capacities): aggregation or disaggregation and other forms of geospatial analysis, (3) simple "tabular" data (e.g., costs of technologies or fuels): varied manual processing such as making assumptions where values are missing or converting currencies.

---

[1] This is akin to the Open Source Definition used by the Open Source Initiative [4], to which all commonly used licences for code conform. The full list of OSI-approved licences is at https://opensource.org/licenses.

[2] http://www.openmod-initiative.org/.

Assumptions made during processing often go unreported and undocumented. As an example, recent work examining the effect of different time resolution reduction methods on model results found large enough effects to qualitatively affect conclusions [10–15]. This suggests that carefully documenting and making processing steps and code open is necessary not just for reproducibility, but also to allow users to assess the impact on results, their interpretation, and resulting policy implications.

### 2.2. Code

Models are idealised representations of real systems built to perform a specific analysis or answer a specific question, and so usually include code (e.g. for reading data, constructing and solving equations) and data (e.g. technology costs). Implementing a model is to turn its conceptual components (such as the equations describing an energy system and the accompanying data parameters) into a computer program, for example, using a mathematical programming language. We distinguish *models* from *frameworks*. The latter are programs which are later populated with data to produce a model. They may contain structures designed to provide reusable functionality when building models (e.g., a general set of functions in a given programming language to read, process and analyse data). Models include data and assumptions, and are usually specific to one situation. The boundaries between model and framework are not always clear, but being aware of them helps when making decisions such as the choice of licence, because licensing code and data requires different considerations [16].

Model implementations vary greatly in complexity. Most simply, models can consist of a collection of spreadsheets. More complex ones might use commercial off-the-shelf tools such as Plexos, a mathematical programming language such as GAMS or GNU MathProg, or a general-purpose programming language such as Java, C++ or Python. Using a mathematical programming language keeps the model code at a higher level and focused on the actual mathematical model. On the other hand, building models directly with general-purpose programming languages allows common components of model implementation to be identified and extracted more easily, and made available for other models to use (i.e., as a framework).

## 3. Key considerations when going open

The multitude of choices and terminology a researcher faces when deciding how to open up code and data can be overwhelming. Based on our collective experience, we provide practical guidance covering the key considerations that arise during the process.

### 3.1. Who owns intellectual property

When making any code or data available, the first important step is to establish who owns the relevant intellectual property, that is, who holds copyright: the researcher or their institution. In most cases where an employment contract is involved, the employer will own the intellectual property rights. Some institutions may automatically grant their researchers the right to open-source research software or have a fast-track process to grant approval for open-source release. Nevertheless, researchers should always confirm ownership regulations with their institutional technology transfer office or legal department [17]. "Provenance" is the history of a codebase and its contributions. Unless provenance can be conclusively determined it cannot be certain that code released under a specific licence is unencumbered by conflicting intellectual property ownership.

Proactively asking contributors to confirm their right to contribute to an open-source project resolves potential legal concerns. For example, as of July 2017, the GitHub terms of service include a clause[3] stating that by making an original contribution to a repository the user agrees to license that contribution under the licence specified in the repository, and that they are legally able to do so (i.e. they own intellectual property for their contribution). Another solution is to include a "certificate" alongside the licensed project that lists contributors including, where applicable, employers or clients that may own rights. Contributors can then be asked for written permission for their contributions where necessary [18], or requested to sign a contribution licence agreement (CLA), although the latter poses an additional barrier to contribution [19]. Such intellectual property questions are not specific to energy modelling, but apply equally to all research software and data.

### 3.2. What and how much to publish

Every bit of information can be supportive when researchers try to reproduce or reuse the work of others. While true reproducibility requires complete openness, it is still valuable to open only parts of the model, data or data processing steps. Several tools exist to support the creation of reproducible research, like entire workflow systems [20], tools to track provenance [21], and more recently containerisation [22]. Research in this field is ongoing [23]. While complete and long-term reproducibility remains difficult and comprises multiple aspects [24], researchers should not shy away from sharing code [25], even if they believe it is not yet comprehensive enough to result in fully replicable science [26].

When sharing either code or data, appropriate documentation for the target audience is important [27]. With code, for example, will users have a graphical interface, or will they use the software as a library included in their own project? Is it desirable that users become collaborators that could extend and improve code? For possible collaborators, the internals of the code and/or the application programming interfaces (APIs) should be documented. In general, best practices from software engineering should be applied where possible [28]. Automated tests can provide a formal specification of the project, as they ensure certain features continue working after changes to the code. Code review can improve code readability and quality [29], and can be used efficiently in a scientific context [30]. Guidelines for researchers new to software development are available in Refs. [31,32]. With data, it is primarily important to document where it has come from, what units and conventions are involved, and where the key uncertainties lie. The use of a standard metadata format, such as the Data Package standard [33], can help to ensure consistent and complete documentation. With full models, beyond documenting their code and data, even just documenting how the model was applied in specific studies can be of help to potential users.

Finally, at what point during a project should code or data be published? A good point to release is when studies relying on the data or code are submitted for review. For continuously developed software under an open license, there is no reason not to develop in the open by keeping code under development on a code hosting platform (see section 3.5) from which released versions are tagged or otherwise archived.

---

[3] GitHub Terms of Service, Contributions Under Repository License: https://help.github.com/articles/github-terms-of-service/#6-contributions-under-repository-license.

Table 1
Common licenses for code and data.

| Applicability | Type | Common licenses |
| --- | --- | --- |
| Code | Copyleft | GNU GPL, LGPL, AGPL, Eclipse, Mozilla |
| Code | Permissive | BSD, MIT, Apache |
| Data | Copyleft | Creative Commons BY-SA, Open Data Commons ODbL |
| Data | Permissive | Creative Commons BY |

### 3.3. Which licence to choose

To reiterate: when code or data are made available it is important to clarify the legal terms under which they are published. Applying a well-known licence is the easiest and preferred approach. It ensures interoperability between software projects and makes it straightforward for users and possible contributors to immediately understand the terms [34]. It is important to note that intellectual property rights and copyright protection always apply. Without a licence, code cannot be legally used [35,36], while the legal context under which data can be used remains unclear for potential users and contributors.

Two key considerations influence the choice of licence. First, different licences are more applicable depending on whether the material to be licensed is code or data. Second is the choice between two types of licences, often grouped into permissive and copyleft licences. Permissive licences generally allow all re-use, including integration into closed-source projects, without requiring improvements to the code to be released at all. In contrast, copyleft licences require that derivative work is shared under the same or a compatible licence. Table 1 gives an overview of common licenses for code and data.

In the case of computer code, the most common permissive licences are the BSD, MIT and Apache 2.0 licences. The most common copyleft licence is the GNU GPL and related licences (such as the LGPL, AGPL) [17,37]. In the case of data and other content, common permissive licences are Creative Commons Attribution licences (CC BY) and common copyleft licences include other Creative Commons licences such as Creative Commons Attribution-ShareAlike (CC BY-SA) or the Open Data Commons Open Database License (ODbL). When licensing code, it is advisable to use licenses specifically developed for code, rather than other licences such as Creative Commons. The reason is that software-focused licences cover technical issues such as linking different pieces of software which do not arise with data and other content.

While some licences (like Creative Commons NonCommercial licences) specifically forbid commercial re-use of data, it is important to be aware that none of the commonly used licences for code prevent commercial use. They only specify whether users (both commercial and non-commercial) must make available their derivative works under the same licence or not. In addition, the definition of "commercial" as opposed to "non-commercial" can be difficult in practice [38]. For instance, contract research may be considered a commercial activity, even if conducted by a university.

The choice of permissive versus copyleft is more intricate. Permissive licences place few restrictions on users and thus make it more likely that code or data are re-used. Copyleft licences, especially the GPL, require that any derivative work is shared under a compatible licence. They are underpinned by a belief in the importance for all code and data to be open, but in practice can restrict the potential set of users: for example, the code licensed under the GPL licence cannot be integrated into permissively licensed code due to conflicting licensing conditions. Further guidance on the licence choice is given in Refs. [17,36,39].

While copyleft licences stipulate that if a derivative work is shared, it must be under the same licence, neither permissive nor copyleft licences require that derivative versions be shared at all. In other words, it is legally possible to publish results obtained with a modified version of a GPL-licensed model, without making available the modified model code. To address this, DeCarolis et al. [40] advocated a new licence with provisions to enforce public release on formal publication of results (e.g. in a journal). However, the existing range of common open-source licenses covers a sufficient range of conditions, and for compatibility reasons, the energy community should use existing licenses as much as possible [34]. Instead of solving this issue via new licenses, journal policies could require public availability of code and data on which submitted work is based.

### 3.4. Which modelling tools or language to choose

The type of modelling tool or programming language used to build a model defines, and to some extent limits, both the capabilities of the model and the degree to which it can easily be made open. The choice should also be considered in light of the user group being targeted [41]. Choosing a specific (open-source) language and licence may be a reason to start building a model from scratch. Open languages, which include most common programming languages, guarantee that there are no financial barriers for new users. They are also often platform-independent, and can be run in containerised environments. On the other hand, commercial tools can offer improved performance [42], capabilities or usability benefits. The barrier to entry for non-programmers is also lower than with fully-featured programming languages. If a model requires a commercial tool such as a proprietary solver to run, it is still possible to provide the code or model definition under an open licence. This does not avoid financial barriers for users, but the model itself can be scrutinised, unless the closed code itself contains implicit or explicit model assumptions [43].

The energy modelling field relies on a wide range of commercial tools, but this does not prevent open release of models and data. Microsoft Excel is widely used, particularly in commercial settings, to develop models ranging from simple spreadsheets to fully packaged applications and web services [44,45]. Its lack of version control and of substantive collaborative features makes it more difficult to use in common open workflows, but downloadable spreadsheets have the advantage of a low entry barrier for users, and can still be released under an open license.

### 3.5. How to distribute code and data

Ways of publishing open code and data range from compressed archives on personal or institutional websites to using code hosting platforms. The target audience and the intention behind open release will influence the choice. A compressed archive is sufficient, and possibly preferable, if minimum maintenance and no active participation is desired (for example, to release an archive of data or code alongside a specific publication). A platform makes sense if further development of code and data alongside contribution from others is envisaged. Platforms like GitHub or GitLab provide issue trackers for feature requests and bug reports, easy ways to contribute and review code, and hosting for wikis and documentation. GitHub in particular has emerged as a standard code sharing platform for many communities [46]. By providing a surface with minimum friction for re-use and collaboration, such platforms are arguably the most appropriate choice for code.

Similar choices have to be made for publishing data. The "good enough practice" described in Ref. [31] includes a description for what kinds of data version control is useful. A comprehensive overview of recommended online repositories for data is provided by Nature Scientific Data[4] For examples of platforms for energy-related data see Refs. [47,48], and for a more general guide on scientific digital data storage, see Ref. [49]. Irrespective of repository choice, a DOI (Digital Object Identifier) makes code and data citable, even if it is not associated with a publication. Online archives such as Zenodo[5] or the Open Science Framework[6] can mint DOIs. For code hosted on GitHub, a new Zenodo DOI can be automatically generated for each release.

### 3.6. How to provide support and build a community

Many open projects do not stop at providing access to code and data, but also offer structures to interact with the developers and support new users. This enables a community to form around a project, which can assist in further development, identifying bugs or other issues and helping people starting out with the model. The nucleus of a community can be formed through code hosting platforms with their issue trackers and wikis. For more on community-building, see Ref. [50].

A common concern is that by providing code and data openly, authors will be flooded with support requests, but that is unfounded in our experience. Conversely, the benefits even for an individual researcher − such as increased citations, requests for contract research and collaborations greatly outweigh the costs. Some projects also go beyond documenting the functionality of their model and put work into creating tutorials and simple examples for beginners; examples include Calliope,[7] OSeMOSYS,[8] PyPSA[9] and oemof[10]. Simple examples allow potential new users to quickly assess the functionality and usability of the model. A step further is to offer a forum or mailing list for people to ask questions, request new features and get update announcements (also see the four above projects for different approaches to doing this). This is common practice for large open software projects. While this may seem to create more time demands for developers and public pressure to respond to support requests, it can also help developers to see how people are using their software or data, may facilitate peer-to-peer support, and may lead to improved documentation and tutorials.

### 4. Examples

The following section uses four examples to illustrate the choices discussed above. First, the difference between starting a new open-source project versus opening up an existing closed model, and the implications of choosing a licence. Second, the additional complexities when licensing data, in particular where mixed sources are involved. Third, a case study of cross-pollination amongst power system tools written in Python. Finally, lessons learned on how to build a community around an energy modelling project. The projects mentioned in these examples are summarized in Table 2.

### 4.1. Starting open or closed and choosing a licence

Creating a new codebase with an open licence intuitively seems easier than retroactively opening a closed-source project. When starting open, all decisions can be made accordingly: for example, choosing third-party code and data based on licence compatibility. A recent flourishing of energy models and frameworks have begun as open-source projects from the outset, for example, OSeMOSYS, oemof, SciGRID, PyPSA, TEMOA, and Calliope (see Table 2). Opening up formerly closed models can be challenging, since it is possible that parts of the code cannot be released (for example, because the copyright holder of the code will not permit it or the provenance cannot be reconstructed). Such parts must be substituted or rewritten, which may not be practical.

Even re-licensing existing open-source code after its release can be confusing and problematic to users. It also requires agreement of all copyright holders, which after several years of development may well include multiple institutions and individuals. Models like SciGRID and Calliope were therefore released under a permissive licence from the start (the Apache 2.0 licence). This permits the inclusion of all or parts of the code into closed-source projects as long as the original copyright notice is included. Other models, like PyPSA or oemof, are licensed under the GPL. The GPLv3 is not bidirectionally compatible with the Apache 2.0 licence, for instance, and if GPLv3 code were added to a project licensed under a permissive licence such as Apache 2.0, the resulting combined codebase would need to be relicensed under the GPLv3. Furthermore, the GPL (both version 2 and 3) expressly forbid distributing software that links to non-GPL-compatible libraries. This can limit re-use options for GPL-licensed code. The PyPSA developers therefore later decided to release part of their code that could be relevant for other projects under the Apache 2.0 licence, which was relatively straightforward as all three developers were working at the same institution at that point.

Besides technical and legal difficulties, moving from closed to open can hit other stumbling blocks. The UK TIMES Model (UKTM) started closed-source and was scheduled to be released with an open licence in 2015 [61]. However, the intended release date has been successively pushed back, and is now anticipated in 2018. Part of the reason given is the additional effort required to meet government guidelines for presentation, documentation and plausibility of models [62]. The use of UKTM within policymaking also creates obstacles to openness. The UK government used it for modelling its response to the Carbon Budgets, the UK's official pathway for decarbonisation, and did not want the model to be openly and freely available whilst conducting this work [63].

A possible way to overcome this issue of culture can be to open things up incrementally. This is particularly pertinent with models that contain large amount of data, not just code. For example, the UK TIMES Model features around 1600 technologies, each of which requires more than a dozen parameters. Unless these data were sourced very carefully, with provenance and attribution established in well-organised metadata associated with the data, opening up such a large model alongside its data might result in almost insurmountable challenges. Improved procedures for attribution, referencing, version control and documentation with such a large catalogue is necessary, in particular since typically, civil servants working with it are typically not trained scientists or data curators [63].

---

[4] https://www.nature.com/sdata/policies/repositories.
[5] https://www.zenodo.org/.
[6] https://osf.io/.
[7] https://calliope.readthedocs.io/en/stable/user/tutorial.html.
[8] http://www.osemosys.org/get-started.html.
[9] https://pypsa.org/examples/.
[10] https://oemof.readthedocs.io/en/stable/getting_started.html#examples.
[11] http://wiki.openmod-initiative.org/.
[12] Main project website with access to code and documentation, or GitHub repository where no separate project website is available.
[13] Release 2001, but licensed only from 2017 onward.

## 4.2. Aggregating and licensing data

Licensing data which was not open from the start can be similarly complex. The Open Power System Data (OPSD) project [47] is a community effort to collect, aggregate and publish data from 100 + sources, most of which had not considered open licensing before. In the European Union, structured data (i.e., databases) can be protected under two different intellectual property right regimes: copyright and the sui generis "database" right. The OPSD project's legal assessment indicated that much of the collected data was likely protected under the sui generis right. Only with all rightholders agreeing to license their data openly could the OPSD database as a whole have been licensed openly to users — but as of 2017 only about half of the rightholders had agreed to open licensing. Reasons ranged from an unclear legal situation on who the data's rightholder actually is, lacking awareness that simply putting data online does not constitute a release into public domain, unwillingness to skate on thin legal ice where rightholders had themselves originally collected data from multiple sources, and finally, the desire not to compromise the ability to sell data.

Indeed, the question of licensing data can veer into the discussion of viable open-source business models. For example, Renewables.ninja [56,57] allows users to run global simulations of hourly power output from wind and solar farms via a web application[14]. Data are made available under a non-commercial Creative Commons license free for academics and not-for-profit organisations. This means that businesses are required to pay for commercial use and thus contribute towards the platform's substantial operating costs. This involves a trade-off between open ideals and financial realities, but also comes with penalties. For example, non-commercial licenses are forbidden from Nature's Scientific Data journal, which explicitly recommends the Creative Commons CC0 waiver [64]. The OpenEnergy Database (oedb), an in-development project to provide a repository for energy model results and input data [48], is also attempting to ensure clarity on licenses by forcing users to supply a license for all uploaded data and implementing automated license compatibility checks on database queries.

## 4.3. Cross-pollination: PYPOWER and its descendants

The range of power system tools written in the Python programming language illustrates the downsides of uncoordinated development: software projects that were started for specific research questions, abandoned, and then further developed by other groups. One of the first open-source power system modelling tool in Python was PYPOWER, a translation of Matlab-based Matpower [65] into Python, developed by Richard Lincoln from 2009 onwards for his doctoral thesis in 2011 [54]. Active development ceased in 2014, although as of 2017, bug fixes from third-party developers are still being merged into the code base.

In 2016, three independent Python power system tools were announced, each of which continued developing PYPOWER towards a different direction. PyPSA [50] re-wrote the code base from scratch and added multi-period optimisation; pandapower [53] built on PYPOWER to further develop the modelling of distribution networks; GridCal [51] added new algorithms for power flow and offered a graphical user interface. While it may seem counter-productive to have so many overlapping software projects in Python, because of the open software licensing, it has been possible to share code and ideas between the projects. PyPSA borrowed some data structures from pandapower; PyPSA's handling of disconnected networks inspired pandapower's, and PyPSA is planning to implement some of GridCal's algorithmic work. Therefore, despite initial fragmentation, users benefit from this pooling of functionality — which is made possible through open licensing.

This example shows the emergence of community efforts through different research groups developing closely related projects, then realising they can benefit from each other's work. It also highlights the issue of projects with no active contribution management — after branching off into separate projects, reunification can be difficult. This implicit community growth therefore has advantages and disadvantages by contrast with the next examples, which include the explicit, focused and labour-intensive effort to create a community around an energy modelling framework.

## 4.4. Community building: OSeMOSYS and oemof

Building and maintaining a community of users and contributors is not straightforward and requires careful consideration. Issues include incentivising users to contribute, attracting new contributors and streamlining the process of engaging with the community. Without focused personal involvement by at least one core developer or community organiser, we have found that this is difficult to achieve.

A good example is OSeMOSYS, a framework for long-term energy system planning optimisation models implemented in GNU MathProg and released in 2009 [7]. The core developers at KTH Stockholm have invested significant time and effort to build a user community. The community can be broadly divided into two types of users: academic researchers (including students) and policy-makers, particularly in developing countries. To reach the first group, the OSeMOSYS developers hold regular side events at conferences such as the International Energy Workshop (IEW). These reach academics looking for open tools to teach energy system modelling, or for building new models for research purposes. For students, OSeMOSYS is useful as a 'gateway' modelling tool — its code is relatively straightforward and easy to customise.

To reach policymakers, the OSeMOSYS team developed a close relationship with two United Nations agencies, UNDESA and UNDP, to help drive adoption of OSeMOSYS by national governments for their energy sector planning. Notable examples of this include South Africa, Cyprus and Bolivia. Open models allow countries with a limited budget for energy policymaking to build their own domestic modelling capabilities. Through such collaboration, the developers engage directly with policy makers. Finally, OSeMOSYS sends out a monthly newsletter with feature additions, upcoming events, and the latest publications.

A slightly different approach was taken by the oemof project [52]. It is a general framework in Python inspired by and based on three earlier models. First, a model of the power system in Germany and neighbouring countries, renpass [58], written in R, as well as the Matlab-based MRESOM [66] and Python-based pahesmf [67]. Oemof, based on these three projects, included some of the original as well as new developers. The key focus was to first establish a working inter-institutional development process by following best practice from professional software development as much as possible, and only later shift to developing an active user community. Ultimately, renpass was then reimplemented using the oemof framework it inspired, as renpassGIS [59]. An initial problem-specific model leading to the development of a more general framework makes sense: indeed another framework, the Calliope project [11], also evolved from an earlier model written in GAMS [68], but in that case, the original inspiration was never re-implemented.

---

[14] https://renewables.ninja.
[15] http://www.openmod-initiative.org/.

Table 2
Overview and classification of the discussed models and frameworks. Type is classified by F (framework), M (model), or D (data project). This list is non-exhaustive: consult other sources such as the Open Energy Modelling Initiative's wiki[11] for a more complete list of open modelling projects.

| Name | Type | Description | Language | Licence | Release | Ref. | Website[12] |
|---|---|---|---|---|---|---|---|
| Balmorel | F | Energy system model framework | GAMS | ISC | 2001[13] | [5] | Link |
| Calliope | F | Energy system model framework | Python | Apache 2.0 | 2013 | [11] | Link |
| deeco | F | Operational optimisation model framework | C++ | GPLv2 | 2004 | [6] | Link |
| GridCal | F | Power system model framework based on PYPOWER | Python | GPLv3 | 2016 | [51] | Link |
| oemof | F | Energy system model framework | Python | GPLv3 | 2015 | [52] | Link |
| oedb | D | Database for energy model input and output data | Python | Data: various Code: AGPLv3 | 2016 | [48] | Link |
| OPSD | D | Power system data platform | PHP | Data: various Code: MIT | 2016 | [47] | Link |
| OSeMOSYS | F | Energy system model framework | MathProg, GAMS, Python | Apache 2.0 | 2009 | [7] | Link |
| pandapower | F | Power system model framework based on PYPOWER | Python | BSD 3-clause | 2016 | [53] | Link |
| PYPOWER | F | Power system model framework | Python | BSD 3-clause | 2011 | [54] | Link |
| PyPSA | F | Power system model framework inspired by PYPOWER | Python | GPLv3 | 2016 | [55] | Link |
| Renewables.ninja | D | Platform for PV and wind generation data and models | Python, R | Data: CC BY-NC 4.0 Code: MIT | 2015 | [56,57] | Link |
| renpass | M | Power system simulation model (Germany and neighbouring countries) | R | GPLv3 | 2014 | [58] | Link |
| renpassGIS | M | German electricity market model | Python | GPLv3 | 2016 | [59] | Link |
| SciGRID | M | Transmission grid model | Python | Apache 2.0 | 2015 | [60] | Link |
| TEMOA | F | Energy system model framework | Python | GPLv2 | 2010 | [8] | Link |
| UKTM | M | UK energy system model based on TIMES | GAMS | Unreleased | – | [61] | – |

## 5. Discussion and conclusion

We are convinced that open energy system modelling – using open data in open-source models to produce open data – comes with many benefits: it increases the quality of research, reduces duplication of work, increases credibility and legitimacy, provides transparency to the policy discourse and makes high-quality data and planning tools accessible to researchers and government agencies without the funds for commercial options. There are also private benefits for individual researchers such as increased citations and future contract research.

Based on our own practical experience we have compiled strategies and lessons learned for researchers wanting to open their modelling black boxes. These include ensuring consent from all intellectual property holders, identifying those parts of the work that can be published, choosing an appropriate licence, considering the programming languages and frameworks to use, identifying an appropriate distribution channel and building a community of users and contributors.

That is not to say that the traditional tools of academic knowledge dissemination are not still useful: researchers should use existing fora such as academic conferences and workshops to engage with the community and discuss methodology, explain models and data accurately in publications, favour concise approaches over complicated ones to answer specific research questions and provide accessible and up-to-date documentation for their models and data.

The decisions we lay out above are predominantly those taken by individual researchers and their research groups. However, there are wider issues for which responsibility lies with the broader institutions which employ and fund these individuals. Three are of particular importance: first, data relevant for energy system research is often provided by institutions such as statistical offices, government agencies, or transmission system operators. These institutions should ensure that open licences are provided from the outset, which sadly today is still the exception rather than the rule. Second, research funding agencies have the power to change the current incentive structure by including open data and open-source requirements in their calls and contract research. Finally, there is still a lack of recognition for software and data development in the academic assessment system. Either they should be recognised alongside paper writing, or it should be easier to write journal papers on software and data. Ultimately, to uphold the trust and equality of academic research, we believe that the methods and results from research funded by public money should be openly available to the public.

## Acknowledgements

Stefan Pfenninger acknowledges support from the European Research Council via grant StG 2012-313553. This paper was made possible by collaboration within the Open Energy Modelling Initiative. We invite readers to join the initiative's email list, forum and regular workshops.[15]